\documentclass[reprint,amsmath,amssymb,amsthm,aps]{revtex4-2}
\usepackage{graphicx}
\usepackage{dcolumn}
\usepackage{bm}

\begin{document}
\preprint{}

\title{Nonlinear Schr\"{o}dinger Equation Solitons on Quantum Droplets}
\author{A. S. Carstea}
\email{acarst@theory.nipne.ro}
\affiliation{Department of Theoretical Physics \\
National Institute of Physics and Nuclear Engineering \\
Bucharest-M\u{a}gurele 077125, Romania}
\author{A. Ludu}
\email{ludua@erau.edu}
\affiliation{Department of Mathematics \\
Embry-Riddle Aeronautical University, Daytona Beach, FL 32114 USA}
\date{\today}

\begin{abstract}
Irrotational flow of a spherical thin liquid layer surrounding a rigid core is described using the defocusing nonlinear Schr\"odinger equation. Accordingly, azimuthal moving nonlinear waves are modeled by periodic dark solitons expressed by elliptic functions. In the quantum regime the algebraic Bethe ansatz is used in order to capture the energy levels of such motions, which we expect to be relevant for the dynamics of the nuclear clusters in deformed heavy nuclei surface modeled by quantum liquid drops. In order to validate the model we match our theoretical energy spectra with experimental results on energy, angular momentum and parity for alpha particle clustering nuclei.  
\end{abstract}

\maketitle

\section{Introduction}
\label{sec1}

Solitons are stable localized wave packets that can propagate long distance in dispersive media without changing their shapes. Following the discovery of solitons by Russell \cite{russell}, a large number of similar particle-like nonlinear localized waves, pulses and finite-gap potentials were identified and discovered, influencing the development of almost all traditional areas of science, and also shaping modern fields of research \cite{solib}. Solitons are studied in a wide range of scales from cosmology and dark matter \cite{cosmology,darkmatter} to quantum scale \cite{fieldtheory} and new states of matter \cite{MatterWaves,BEC1,BEC2}, and they occur in a broad spectrum of systems, from low temperature \cite{LowTemp} to nonlinear biological or social systems \cite{biology1}. Soliton theory initiated major developments in optical communication \cite{OptSolitFiberLasers}, especially by revealing universality properties of several nonlinear phenomena like rogue waves \cite{rogue2,rogue3}, in anomalous materials  \cite{AnomalousTranspRelaxation}, or in the collective dynamics of large random ensembles (soliton gas, soliton rain) \cite{SolitonGas,rain}. Solitons helped the development of new applications in  technology: soliton computing \cite{QComputersSolitonComputing}, machine learning
\cite{MachineLearning2}, or non-Hermitian optics \cite{NonHermitianPhotonics}. At present, the long-range soliton stability is so well understood that ordered set of solitons are used to carry out the transmission of information in fiber optics communication links \cite{PhotonicCrystalFibers}. 

Equivalently, the question of long life-time solitons confined in a compact region  \cite{prl,book} represents a subject of active research in ocean wave dynamics.  
Solitons are present not only in long and narrow geometries such as channels, fiber optics, electric lines, or nerves, but they were also found as soliton gas or periodic waves in compact regions \cite{book}, and in bounded nonlinear optics systems \cite{SolitonGas, rogue2}. A rain of soliton pulses, triggered by a noisy background, can start flowing inside a finite fiber laser cavity, together with its condensed phase \cite{rain}. Such trains of bound solitons (soliton molecules \cite{mole}) can also travel at constant angular frequency  through circular fiber rings \cite{ring}. It was also possible to generate multi-soliton rotating clusters and quasi-polygonal stable soliton clusters in bulk nonlinear optical media \cite{goza,*geza}. 

Rotating solitons/solitary waves can occur in microscopic systems. Such excitations are theoretically obtained in the quantum Hall effect of 2D electron drops \cite{electrons}, or in Bose-Einstein condensates \cite{bec}, and they were measured in superfluid helium rotating vortices \cite{qvor}. 

At lab scale, the formation of periodic nonlinear waves, or cnoidal waves for  Korteweg-de Vries models (KdV), on closed and bounded systems was detected and the results were matched with theoretical calculations in low temperature interfacial systems \cite{qvor,myleiden}, in confined rotating flows \cite{hamid}, and along circular chains of magnetic pendulums \cite{magn}. Experiments demonstrate the formation of rotating hollow polygons in 2D fluids, within good match with theoretical models of cnoidal waves  \cite{cnoidalleiden,myleiden,dropsmagnlevitated,leidentorus,bohr,rot,fr,rot2,flu}.

Cnoidal patterns and solitary waves at large scales were observed as vortex waves \cite{vortexfriedland},  and as rotating polygons in the hurricane eye wall \cite{hurric}, as well as in the case of Saturn's North Pole hexagon \cite{saturn}. Numerical simulations for the azimuthal nonlinear surface waves on neutron stars surrounding a rigid core generate localized, shock-type dispersionless solutions \cite{neutronstar}.

The formation of solitons on spherical surfaces was considered as a possible explanation of large amplitude collective modes of excitation on nuclear surfaces  in the liquid drop model for  cluster radioactivity  \cite{first}, or as shape solitons on the surface of liquid drops \cite{prl}. Nonlinear models with soliton solutions offer possible explanations for the emergence of such \textit{rotons} as coherent states in nuclear systems \cite{coherentalpha}, in $\alpha$ particles collision with medium-heavy nuclei \cite{finland,neon,jpg}, in nuclear fission \cite{gherghescu}, and in cnoidal excitations of  Fermi-Pasta-Ulam rings \cite{solitonsring}. 

These results suggest that some dynamical systems can have collective localized stable excitations  in compact or bounded geometries.  Given the observed similarity between  such rotating solitary waves within various ranges of physical scales (from nuclei to neutron stars) there may be a possibility of  manifestation of signatures of universality.

In this paper we show that for a spherical thin liquid droplet surrounding a rigid core one can develop an asymptotic procedure which gives the evolution of periodic envelope solitons in the azimuthal direction (the spherical $\varphi$ coordinate). The variation in the polar coordinate $\theta$ is considered to be very slow (more precisely this approximation is valid not very close to the spherical poles). The asymptotic (related to the thickness of the spherical fluid layer) of Laplace equations and kinematic boundary condition transforms the linearized spherical Euler equation into a {\it nonlinear} one, supporting plane wave solutions with a Boussinesq-type dispersion relation. In the full nonlinear Euler equation we assume that in stretched space-time scale, a slow modulation of the plane wave occurs and accordingly, a defocusing nonlinear Schr\"{o}dinger equation is obtained. Periodic dark solitons solutions expressed by elliptic functions are described. In a sense, this paper is a continuation of \cite{prl} where cnoidal KdV 1-phase solutions were founded.  In section \ref{sec.quant} the last part we analyze the quantum dynamics of such system using algebraic Bethe ansatz, a well known procedure for the defocusing nonlinear Schr\"odinger equation. We believe that this fact to be relevant in the study of collective excitations of the surface of heavy nuclei in exotic radioactivity processes  \cite{first}. In the last section we match our theoretical energy spectra with experimental results on energy, angular momentum and parity for alpha particle clustering nuclei for atomic masses ranging from $20$ to $212$.

\section{General Derivation of the Nonlinear Schr\"{o}dinger Equation}
\label{sec2}

Solitons represent fundamental nonlinear modes of physical systems described by a special class of wave equations of an integrable nature. These equations,
like the KdV equation or the nonlinear Schr\"{o}dinger equation (NLS), are of significant physical importance since they describe at the leading order the behavior of many systems in various fields of physics 

Our model is an ideal spherical liquid layer exhibiting irrotational flow. The inner surface is bounded by a rigid core of radius $R_0 -h$, and the variable outer surface $\Sigma$ is paramaterized by spherical coordinates $r=R_0(1+\xi(\theta,\phi,t))$.
We further assume that traveling perturbations  will be slowly varying in $\theta$ and the fast dynamics is happening in the $\phi$ direction and we separate $\xi(\theta,\phi,t)=g(\theta)\eta(\phi,t)$, with $g(\theta)$ a slowly varying function.
From the equation of continuity for incompressible fluid $\rho=$const. and irrotational condition we have the Laplace $\Delta\Phi=0$ and Euler equation:
$$
\biggl(\Phi_t+\frac{1}{2}|\nabla\Phi|^2\biggr)_{\Sigma}=-\frac{P}{\rho},
$$
where $P$ is the pressure and $\Phi$ is the velocity potential. The boundary condition on $\Sigma$ 
$$
\frac{dr}{dt}\biggr|_{\Sigma}=\left(\partial_t r+\frac{d\theta}{dt}\partial_{\theta}r+\frac{d\phi}{dt}\partial_{\phi}r\right)_{\Sigma},
$$
can be written in terms of the velocity potential in spherical coordinates:
$$
\Phi_{r}|_{\Sigma}=R_0\left(\xi_t+\frac{\xi_{\theta}}{r^2}\Phi_{\theta}+\frac{\xi_{\phi}}{r^2\sin^2\theta}\Phi_{\phi}\right)_{\Sigma}.
$$
Because our model is a liquid shell we have the inner boundary condition $ v_r=\partial_r\Phi|_{r=R_0-h}=0$.
In order to meet the harmonic condition we expand the flow potential \cite{prl,book}
$$
\Phi=\sum_{n=0}^{\infty}\left(\frac{r-R_0}{R_0}\right)^n f_n(\theta,\phi,t),
$$
where the functions $f_n$ must obey recursion relations obtained form the Laplace equation. Assuming the smallness parameter $h/R_0=\epsilon<<1$ and $(r-R_0)/R_0=\epsilon$ we obtain the following relations in the dominant order from the inner boundary condition 
$$
f_1=2\epsilon f_2,\quad f_2=-\frac{1}{2}(\Delta_{\Omega}f_0+2f_1).
$$
From the free surface boundary condition and slowly variation on $\theta$ we can write
\begin{equation}
f_{0,\phi}=\frac{R_0^2\sin^2\theta\xi\xi_t}{\epsilon\xi_{\phi}}+\mathcal{O}(\xi^2).
\end{equation} 
Also using the expansion of velocity potential we have in the first order
$$
\Phi=f_0+\xi f_1+\mathcal{O}(\xi^2),\quad \Phi_{\phi}=f_{0,\phi}+\mathcal{O}(\xi^2),
$$
$$
v_{\phi}=\frac{\Phi_{\phi}}{r\sin\theta}=\frac{f_{0,\phi}}{R_0\sin\theta}.
$$
Deriving with respect to $\phi$ the Euler equation we get (we neglect the $\theta$ derivatives)
\begin{equation}\label{euler}
 \partial_t(f_{0,\phi}+\xi f_{1,\phi}+...)+\partial_{\phi}(\frac{v_{\phi}^2}{2})=\frac{2\sigma}{\rho R_0}\xi_{\phi}+\frac{\sigma}{\rho R_0}\Delta_{\Omega}\xi_{\phi}+\mathcal{O}(\xi^2),
\end{equation}
we obtain 
$$
\partial_t\left(\frac{R_0^2\sin^2\theta\xi\xi_t}{\epsilon\xi_{\phi}}\right)+\partial_{\phi}\left(\frac{R_0^2\sin^2\theta\xi^2\xi_t^2}{2\epsilon^2\xi_{\phi}^2}\right)-\frac{2\sigma\xi_{\phi}}{\rho R_0}-\frac{\sigma\xi_{\phi\phi\phi}}{\rho R_0\sin^2\theta}=0.
$$
The linearized version of the  Euler equation is given by:
$$
\Phi_t=-\frac{1}{\rho}P,
$$
which further can be written
$$
\partial_t\left(\frac{R_0^2\sin^2\theta\xi\xi_t}{\epsilon\xi_{\phi}}\right)=\frac{2\sigma}{\rho R_0}\xi_{\phi}+\frac{\sigma}{\rho R_0\sin^2\theta}\xi_{\phi\phi\phi}.
$$
This equation admits linear traveling wave solution $\xi=A(\theta)e^{i(k\phi-\omega t)}$+c.c. with the Boussinesq-type dispersion relation
$$
\omega^2=\frac{\epsilon\sigma}{\rho R_0^3\sin^2\theta}\left(2k^2-\frac{k^4}{\sin^2\theta}\right).
$$
In the long-wave limit $k=\epsilon K\sim \mathcal{O}(\epsilon)$ the dispersion relation becomes 
$$
\omega(K)=\frac{\epsilon^{1/2}}{R_0\sin\theta}\sqrt{\frac{2\sigma}{\rho R_0}}\left(\epsilon K-\frac{\epsilon^3 K^3}{4\sin^2\theta}\right)+\mathcal{O}(K^5)
$$
\begin{equation}\label{dis}
\equiv \epsilon v K+\beta \epsilon^{7/2} K^3+\mathcal{O}(K^{11/2}),
\end{equation}
with $v=\epsilon^{1/2}/(R_0\sin\theta)\sqrt{2\sigma/\rho R_0}$ being the phase velocity. It results that  for the monochromatic case this dispersion provides exactly the stretched variables for the KdV equation. Indeed from Eq. (\ref{dis}) we have
$$
\xi=A\exp(i(kx-\omega t))=A\exp [ iK ( \epsilon(\phi-vt)-\epsilon^{7/2}\beta K^3 t-\dots)] ,
$$
and this suggests the variables
$\phi\to\epsilon(\phi-vt), T\sim \epsilon^{7/2} t, \xi\to \epsilon^3 g(\theta)\eta(\phi,T)$. In this new variables one obtains immediately the KdV equation in $\eta(\phi,T)$, which is analyzed extensively in \cite{prl,book}

In the following, one can see that for $\xi(\phi,t)=Ae^{i(k\phi-\omega t)}$ we have $\xi\xi_t/\xi_{\phi}=-(\omega/k) \xi\equiv -v\xi$. 
This nonlinearity produces higher harmonics and weakly modulation of amplitude in the slow variables $\varphi,\tau$ which will de defined next. So we are going to make the following approximation:
$$
\frac{\xi\xi_t}{\xi_{\phi}} \sim \xi(\phi,\theta,t)=\sum_{n=-\infty}^{\infty}\varepsilon^{s_n}Q_n(\varphi,\tau,\theta)e^{in(k\phi-\omega t)}.
$$
Here $\varepsilon$ is a small parameter measuring the weak modulation of the amplitude in a slow space-time scale (different form $\epsilon=h/R_0$) and $s_n$ are some exponents which have to be determined form balance. Now we can define the slow variables
$$
\varphi=\varepsilon\left(\phi-\frac{2\sigma t}{v\rho R_0^2\sin^2\theta}+\frac{3\sigma k^2t}{v\rho R_0^3\sin^4\theta}\right),
$$
$$
\tau=-\varepsilon^2 k t,
$$
and the amplitudes of the expansion:
$$
Q_0(\varphi,\theta,\tau)=\varepsilon^2 g(\theta)V_0(\varphi,\tau),
$$
$$
Q_2(\varphi,\theta,\tau)=\varepsilon^2 g(\theta)V_2(\varphi,\tau),
$$
$$
Q_n(\varphi,\theta,\tau)=\varepsilon^n g(\theta)V_n(\varphi,\tau),\quad n\neq 0,2, \quad V_{-n}=V_n^*.
$$
Introducing these expressions from above in the Euler equation we obtain the following defocusing Nonlinear Schr\"odinger equation with dimensionless terms
\begin{equation}\label{eq.nls3}
i\frac{A(\theta)}{3D(\theta)}\frac{\partial \zeta}{\partial \tau}+\frac{\partial^2\zeta}{\partial\varphi^2}-\frac{C(\theta)^2g(\theta)^2}{18D(\theta)^2}|\zeta|^2\zeta=0,
\end{equation}
where we used the following notations:
$$
\zeta(\varphi,\tau)=\frac{V_1}{k}, \ A(\theta)=\frac{v R_0^2\sin^2\theta}{h},
$$
$$
C(\theta)=\frac{v^2 R_0^4\sin^4\theta}{h^2}, \ D(\theta)=-\frac{\sigma}{\rho R_0\sin^2\theta}.
$$
The physical configuration is give by the parameterization equation
$$
r=R_0(1+g(\theta)(\epsilon k\zeta(\varphi,\tau)e^{i(k\phi+\omega t)}+{\rm c.c.}+$$
$$+\epsilon^2 (\frac{Cg}{6D}\zeta^2-\frac{Cg}{3D}|\zeta|^2 e^{2i(k\phi+\omega t)}+{\rm c.c}))).
$$

\subsection{Dark periodic soliton}

In order to obtain periodic solutions and traveling waves, we consider
$$
\zeta(\varphi,\tau)=f(k\varphi+\omega \tau)e^{i(\lambda\varphi+\Omega\tau)},
$$
and we make the shorthand notations $s=k\varphi+\omega \tau$ and $\eta=\lambda\varphi+\Omega\tau$.By introducing these notations in the NLS Eq. (\ref{eq.nls3}) it results
$$
(18D\lambda^2+6A\Omega)f(s)-6i(6Dk\lambda+A\omega)f'(s)+
$$
\begin{equation}\label{eq7773}
+C^2Dg^2f(s)^3-18Dk^2f''(s)=0.
\end{equation}
By imposing $6Dk\lambda+A\omega=0$ we obtain an equation which can be solved by elliptic functions. To make it simpler we divide by $18Dk^2$ and we find
\begin{equation}\label{43356}
b_0 f(s)+b_1 f(s)^3-f''(s)=0,
\end{equation}
where 
$$
b_0=\left(\frac{\lambda}{k}\right)^2+\frac{A\Omega}{3Dk^2}, \quad b_1=-\frac{C^2g^2}{18k^2},\quad \omega=-\frac{6Dk\lambda}{A}.
$$
The solution of Eq. (\ref{43356}) is
\begin{equation}\label{eqsolu}
f(s)\equiv f(\varphi,\tau)=H{\rm sn}\left(k\sqrt{\frac{C^2g^2}{18k^2(m+1)}}\left(\varphi
-\frac{6D\lambda}{A}\tau\right)\biggl| m\right)
\end{equation}
where $H=i\sqrt{{2b_1 m}/{b_0(m+1)}}$. When $m\to 1$ we obtain the dark line-soliton limit
$$
f(\varphi,\tau)\to i\sqrt{\frac{b_1}{b_0}}{\rm Tanh}\left(\sqrt{\frac{-b_1}{2}}k\left(\varphi-\frac{6D\lambda}{A}\tau\right)\right).
$$
We stress that the solution in Eq. (\ref{eqsolu}) is a particular one. The most general solution has the form 
$$
\zeta(\varphi-\nu\tau)=\sqrt{f(\varphi-\nu\tau)}\exp[ig(\varphi-\nu\tau)],
$$ 
and it can be expressed in terms of the Jacobi $sn$ function and the elliptic integral of the third kind
$$
f(x)=a_1+(a_2-a_1){\rm sn}^2 \biggl(\sqrt{\frac{c(a_3-a_1)}{2}}x \biggl| m \biggr),
$$
$$
g(x)=\frac{\nu x}{2}+\sqrt{\frac{a_2a_3}{a_1 a_3-a_1^2}}\Pi \biggl[1-\frac{a_2}{a_1};{\rm am}\biggl( \sqrt{\frac{c(a_3-a_1)}{2}}x \biggr) \biggl| m  \biggr],
$$
where $a_1,a_2,a_3$ are the roots of the ``potential'' equation related to $f$ 
and $m=\sqrt{(a_2-a_1)/(a_3-a_1)}$, see \cite{stef} for details.

\section{Quantization}
\label{sec.quant}
Our equation Eq. (\ref{eq.nls3}) can be written in a Hamiltonian form:
\begin{equation}\label{ham}
\partial_T\zeta=\frac{\delta}{\delta\zeta^{\dagger}}\int_{0}^{2\pi}\left(|\zeta_{\varphi}|^2+c|\zeta|^4\right)d\varphi,
\end{equation}
where we rescaled time $T=(A/3D)\tau$, and $c=C^2g^2/36D^2$.
The pseudovacuum is $|0>$ and $\zeta(\varphi)|0>=0$. In order to perform the quantization, we discretize the system on a lattice, which means that our system will not be defined any more on the meridian circle of the spheroidal drop, but on a polygon with $M$ sides. Also the evolution variable is the angle $\varphi:=n$ which is increased/decreased by fixed step-angle $h$. Lax operator with $\lambda$ spectral parameter is \cite{bethe}
$$L_n(\lambda)=
\left(
\begin{array}{cc}
1-\frac{i\lambda h}{2}&-ih\sqrt{c}\zeta_n^{\dagger}\\
ih\sqrt{c}\zeta_n&1+\frac{i\lambda h}{2}\\
\end{array}\right)+\mathcal{O}(h^2),
$$
and the quantum operators obey $[\zeta_n,\zeta_m^{\dagger}]=\delta_{nm}/h$, where we consider that $\hbar\equiv 1$.

Because we have periodic boundary condition the Lax operator is transformed to monodromy operator. Namely, by imposing periodicity we have the transition from zero-curvature formulation to the pure Lax formulation, by using the monodromy matrix
$$
T(\lambda)=L_{M}(\lambda)...L_{1}(\lambda)=
\left(
\begin{array}{cc}
A(\lambda)&B(\lambda)\\
C(\lambda)&D(\lambda)\\
\end{array}\right),
$$
where $A,B,C,D$ are operators, and not the coefficients of the initial KdV or NLS equations.
\begin{figure*}
	\centering
	\includegraphics[scale=.8]{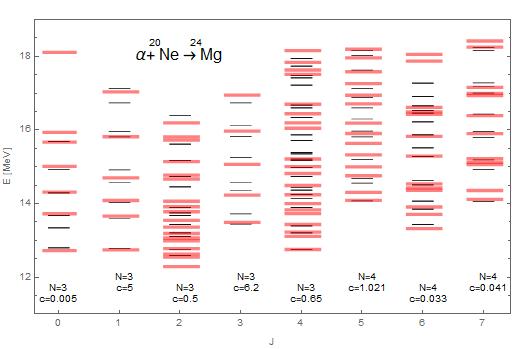}
	\caption{\textit{Black:} Experimental energy spectra, \cite{neon}, of positive- and negative-parity resonant states obtained in the collision of $\alpha-$particles on $^{20}$Ne targets with formation of bound $\alpha-$cluster states in $^{24}$Mg. The spectra are horizontally aligned by angular momentum $J$ from $J=0^{+}$ to $J=7^{-}$. \textit{Red:} The theoretical Bethe spectra Eq. (\ref{eq.bethe1}) are plotted for rapidities $N=3$ and $4$ with the parameter $c$ chosen to provide the best fit with experiments. The odd angular momentum states (labeled with higher placed text in the figure) provide a good fit for larger values of $c$, typically $c>c_{crit.}$, while the best fit for even states occur for relative smaller $c$, shown in the figure under each column.}
	\label{fig1}
\end{figure*}
The evolution of Lax operator can be written either with a new matrix $P$ in the form
$L_t=[P,L]$ or equivalently using the R-matrix formalism, which singles out the Hamiltonian structure. When we quantize, we can write explicitly the commutation relation between elements of monodromy matrix using the so-called RTT-relation
$$
R(\lambda,\mu)(T(\lambda)\otimes T(\mu))=(T(\mu)\otimes T(\lambda))R(\lambda,\mu),
$$
where the matrix $R$ is given by:
$$R=\left(
\begin{array}{cccc}
f(\lambda,\mu)&0&0&0\\
0&g(\lambda,\mu)&1&0\\
0&1&g(\lambda,\mu)&0\\
0&0&0&f(\lambda,\mu)\\
\end{array}\right),
$$
with $f(\lambda,\mu)=1+ic/(\mu-\lambda), g(\lambda,\mu)=ic/(\mu-\lambda)$. Here $2c$ is the $\theta-$dependent coefficient of our NLS Eq. (\ref{eq.nls3}),  $C^2g^2/18D^2$.
The action of elements of the monodromy matrix on the vacuum is 
$$
A(\lambda)|0>=a(\lambda)|0>, D(\lambda)|0>=d(\lambda)|0>,
$$
$$ 
C(\lambda)|0>=0, B(\lambda)|0>={\rm free}.
$$
As a result, one can see that in our case 
$$
a(\lambda)=\prod^M (1-i\lambda h/2)=(1-i\lambda h/2)^M, 
$$
$$
\lim_{M\to\infty}a(\lambda)=e^{-i\lambda Mh/2}, \ 
d(\lambda)=(1+i\lambda h/2)^M, 
$$
$$
\lim_{M\to\infty}(1+i\lambda h/2)^M=e^{i\lambda Mh/2}.
$$
We can further use $Mh\to 2\pi$, since the full periodicity is of $2\pi$ angle.
The quantum states are constructed by applying operator $B(\lambda_i)$ from the monodromy matrix. In the case of $N$ parameters (usually called rapidities) we have:
$$
\Psi(\lambda_1,...,\lambda_N)=\prod_{j=1}^NB(\lambda_j)|0>.
$$
Now imposing that this $\Psi$ must be an eigenvector of the {\it trace} of the monodromy matrix we find the following Bethe equations:
\begin{equation}\label{eqBA}
e^{2\pi  i\lambda_{m}}=\prod_{j=1,j\neq m}^N\left(\frac{\lambda_{m}-\lambda_j+i\frac{C^2g^2}{36D^2}}{\lambda_m-\lambda_j-i\frac{C^2g^2}{36D^2}} \right),
\end{equation}
and the eigenvalue of ${\rm Trace}T(\mu)$ are 
$$
{\rm Trace}T(\mu)\Psi=(A(\mu)+D(\mu))\Psi=\Lambda\Psi,
$$
with
$$
\Lambda=e^{-i\mu\pi}\prod_{j=1}^N f(\mu, \lambda_j)+e^{i\mu\pi}\prod_{j=1}^N f(\lambda_j,\mu),
$$ 
where, as it was shown above, $f(\mu,\lambda)=1+ic/(\lambda-\mu)$. It is easy to note that all the quasi-momenta $\lambda_j$ are dimensionless.
\begin{figure}
	\centering
	\includegraphics[scale=.75]{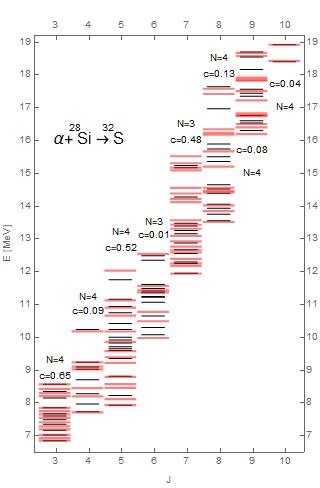}
	\caption{\textit{Black:} Experimental energy spectra, \cite{finland}, of positive- and negative-parity resonant states obtained in the collision of $\alpha-$particles on $^{28}$Si targets with formation of bound $\alpha-$cluster states in $^{32}$S, plotted vs. $J$ angular momentum. \textit{Red:} The theoretical Bethe spectra Eq. (\ref{eq.bethe1}) are plotted for $N=3$ and $4$ with the parameter $c$ chosen to provide the best fit with experiments. The odd angular momentum states are again associated with larger values of $c$, shown in the figure on top of each column.}
	\label{fig2}
\end{figure}
Since it is well known that the trace of the monodromy matrix is nothing but the generating function of conserved integrals of motion (as a power series in $1/\mu$ in our case), we can finally write the eigenvalues of the Hamiltonian
$$
E_N=\sum_{m=1}^N\lambda_m^2,
$$
where $N$ is the number of particles associated with $\Psi$, and $\lambda_m$ are the solutions of the transcendental Bethe equations Eq. (\ref{eqBA}),
%\begin{equation}\label{eqBA}
%e^{2\pi  i\lambda_{m}}=\prod_{j=1,j\neq %m}^N\left(\frac{\lambda_{m}-\lambda_j+i\frac{C^2g^2}{36D^2}}{\lambda_m-\lambda_j-i\frac%{C^2g^2}{36D^2}} \right),
%\end{equation}
so we will have quantum levels parameterized by the polar angle $c=c(\theta)$ with the real values
\begin{equation}\label{eq.bethe1}
E_N (c)=\frac{\hbar^2}{R_0^2}\sum_{j=1}^N\lambda_j^2 (c).
\end{equation}
Eq. (\ref{eqBA}) has only real solutions and they are all periodic of period $1$, so it is enough to consider the solutions $\lambda_j \in [0,1]$.

\section{Discussion}
One can ask what is the role of dark solitons and how their dynamics is seen in the quantum regime. First of all as we have seen we obtained a nonlinear Schrodinger equation 
with defocusing nonlinearity. On the spatial infinite line the soliton solutions are rarefaction (dark) nonlinear waves which are build upon a finite condensate. 
But for the periodic boundary conditions these rarefaction waves are turned into periodic solitons expressed through Jacobi elliptic functions. They describe periodic 
enevelopes of azimutal excitations with various periodicities. In the quantum regime the defocusing nonlinear Schrodinger equation is nothing but interacting delta-Bose gas (with periodic boundary conditions). The algebraic Bethe ansatz
provides a quantisation of the whole dynamical system described by the Hamiltonian Eq. (\ref{ham}) and {\it not} a quantisation of a special classical solution (periodic dark soliton). 
However there is a correspondence between the periodic dark soliton and the expectation value of the density operator on a {\it special} Bethe quantum 
state $<\Psi_0|\zeta^{\dagger}(\varphi,t)\zeta(\varphi,t)|\Psi_0>$ \cite{prljap} constructed using some specific Bethe numbers. The construction is complicated and involves numerical simulations.

\begin{figure}
	\centering
	\includegraphics[scale=.55]{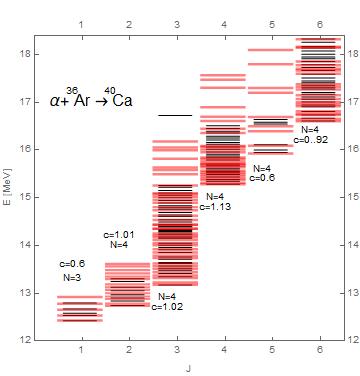}
	\caption{\textit{Black:} Experimental energy spectra, \cite{2011alpha}, of positive- and negative-parity resonant states, plotted vs. $J$, measured during collisions of $\alpha-$particles on $^{36}$Ar targets, with formation of bound $\alpha-$cluster states in $^{40}$Ca. \textit{Red:} The theoretical Bethe spectra Eq. (\ref{eq.bethe1}) for $N=3$ and $4$ and  $c$ values shown in the figure above/below each column.}
	\label{fig3}
\end{figure}
The structure of the energy spectra changes with $N$ and with the parameter $c=C^2 g^2 / 36 D^2$. For any given $N$ its is observed that there is always a region for the parameter $c=c_{crit.}$ around which the energy spectrum becomes very dense. For values $c<c_{crt.}$ the spectral lines are  rather equidistant, while for larger $c>c_{crt}$ the spectrum tends to be quadratic, similar with the spectrum for the rigid rotor. The larger the number of eigenvalues $N$, the smaller the value of $c_{crt.}$ is.  

From the expression of the coefficients $C,D$ from liquid drop model introduced  in \cite{prl} it results that $c\simeq 0$ at $\theta=0,\pi$ as expected since there are no soliton excitation orbiting at the  poles of the droplet. In general, $c$ has its larger values around the equator $\theta \simeq \pi/2$. For certain combinations between the soliton orbital speed $V$, the depth of the shallow layer $h$ and the strength of the surface tension coefficient $\sigma$, the parameter $c(\theta)$ can be very small for all polar angles $\theta$. For example, for fast solitons orbiting a very shallow layer $h/R_0 \ll 1$ and for weak surface tension, one can have very small values of $c$, resulting in weak energy excitation, resulting in a small probability to excite such soliton solutions. For droplets of the size of a medium-heavy nucleus, the parameter $c$ acquires values around $c\sim 10^3$ for almost all values of $\theta$, resulting in larger probability to excite such soliton excitations.  
\begin{figure}
	\centering
	\includegraphics[scale=.85]{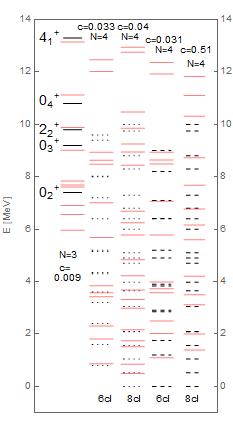}
	\caption{\textit{Solid black:} Experimental energy spectra, angular momenta and parity of  bound $\alpha-$cluster states in $^{12}$C. \textit{Dotted and dashed black:}  theoretical energy excitations calculated for 100 \% and 70\% condensation, respectively  \cite{2018bec}. The Hoyle state is shifted here to $0$ MeV.  Under each of the columns 2-5  we show the number of $\alpha-$clusters considered in the condensation model \cite{2018bec}: $n=6$ clusters for $^{24}$Mg (columns 2, 3) and  $n=8$ clusters for $^{32}$S (columns 4, 5). \textit{Red:} The theoretical Bethe spectra Eq. (\ref{eq.bethe1}) for the best fit with parameters $N, c$, shown on top of each column.}
	\label{fig4}
\end{figure}

Eq. (\ref{eqBA}) can be re-written in the form
\begin{equation}\label{eqBA2}
2 \pi \hat{\lambda}_{j}=2c \sum_{j=1,j\neq N}^{N}\frac{\lambda_j -\lambda_k}{(\lambda_j -\lambda_k)^2 -c^2},
\end{equation}
where the hat symbol represents the equivalence class modulo addition of integers  $n_j \in \mathbb{Z}$. We can evaluate the solutions of this equation for large values of the parameter $c \gg \lambda$. In this case we can make the approximation $|\lambda_j -\lambda_k | \ll c$, and  Eq. (\ref{eqBA2}) becomes a linear system of equations in $\lambda_j$ with the free term given by a column of $N$ arbitrary integers $(n_1, \dots , n_j, \dots, n_N)^{T}$. It is straightforward to calculate the solutions of this linear system
$$
\lambda_j \simeq \frac{\sum_{k=1,k\neq j}^{N}n_k +(-1-\pi c-2N ) n_j}{\pi c(\pi c+N)}\simeq \frac{\hat{1}}{\pi c},
$$
and consequently, in this approximation, the spectrum has a quadratic structure
\begin{equation}\label{eq.bethe2}
E_{N}\simeq \sum_{k=1}^{N} \frac{n_{k}^{2}}{\pi^2 c^2}, \ n_k \in \mathbb{Z}, 
\end{equation}
which is manifested for intermediate values for $c>\lambda$. Nevertheless, since $\lambda\in [0,1]$ there are limitations on the values of the arbitrary integers $n_k$ and in fact this constraints requests $n_k$ to be of order of the integer part of  $(N-1)/2$ meaning that all $n_k$ are constant and the spectrum is actually represented by a constant multiplied by a sum of ones. This observation explains the asymptotic behavior of the spectrum for large $c$ towards a harmonic oscillator spectrum. In fact, in the limit $c\rightarrow \infty$ Eq. (\ref{eqBA}) reduces to $\exp(2 \pi i \lambda_j) =\pm 1$ which generates equidistant energy lines spectrum, since $\hat{\lambda}_j =1/2$.
\begin{figure}
	\centering
	\includegraphics[scale=.85]{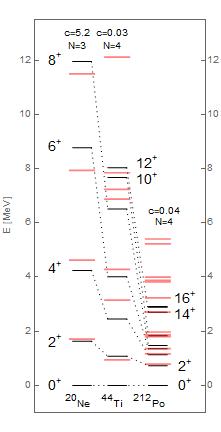}
	\caption{\textit{Solid black:} Energy, angular momentum and parity in three columns representing spectra of bound $\alpha-$cluster states, for $^{20}$Ne, $^{44}$Ti, and $^{212}$Po, respectively \cite{2018NeTi}. \textit{Red:} Best fit for the theoretical Bethe spectra Eq. (\ref{eq.bethe1}) with resulting parameters $N, c$ shown on top of each column.}
	\label{fig5}
\end{figure}
The spectral density of the $\lambda_j$ solutions as a function of $c,N$ can be estimated by introducing a vector nonlinear operator $\mathfrak{O}=(\mathfrak{O}_{i})$,  acting on the $N-$dimensional unit cube of vectors $\vec{\lambda}=(\lambda_1, \dots, \lambda_N)$ in the form
$$
\mathfrak{O}_i(\vec{\lambda})= \sum_{k=1,k\neq i}^{N} \frac{\lambda_i -\lambda_k}{(\lambda_i -\lambda_k)^2-c^2}.
$$
With this operator the Bethe ansatz Eq. (\ref{eqBA}) becomes a fixed point equation for this operator, and while acting on the vectors $\vec{\lambda}$, $\mathfrak{O}$ is a contraction, so it has a unique fixed point, and thus the eigenvector space reduces to one vector. For this case the energy spectrum has one spectral line only. Consequently, dense energy spectra are in the regions where this operator is not a contraction. For such regions inside the unit cube and for the corresponding values of $c$, the spectrum becomes rich in spectral lines.
Obviously, when $|c|<1$ there are values for $\vec{\lambda}$ where the denominators in Eq. (\ref{eqBA2}) approach zero, so the operator is described by Lipschitz discontinues function and thus $\mathfrak{O}$ 
cannot be a contraction. 
For these regions the spectrum becomes denser, as one can easily verify by numerical calculations  in the region $c\simeq1$. However we have to underline that in the limit of $c\to 0$ we have 
free bosons while in the limit $c\to\infty$ the defocusing intercation is so huge and we have free fermions (inasmuch as the Bethe state obeys the Pauli principle). The Pauli principle for Bethe states shows that the nonlinear excitations of the quantum liquid drop model are purely fermionic.

\section{Comparison with nuclear experimental data} 

In order to validate the physical relevance of the  quantum nonlinear liquid drop representation introduced here, we compare the energy levels predicted by our model with experimentally measured resonant lines for alpha clustering  nuclei \cite{neon,finland,gherghescu,jpg,1997Abbodann,2010clusters,2011alpha,2018alpha,2018bec,2018TeXe,2018NeTi}. 
Numerous studies show that alpha clustering occurs from light and
medium-mass elements to heavy and superheavy elements. Various phenomenological and
microscopic models have been proposed in literature to describe various aspects of alpha
clustering, \cite{1997Abbodann,2010clusters,2011alpha,2018alpha,2018bec,2018TeXe,2018NeTi}. Among them, the large amplitude nonlinear collective model, \cite{first,prl,finland,x21,neon,gherghescu,coherentalpha,jpg,book},  are of special interest to the present work.

In the following, we mention four classes of experimental observations and the associated  theoretical questions, pointing the  interest towards using nonlinear collective models. These type of models can relate the features of super-deformed nuclei, cluster radioactivity, quasimolecular structures, or alpha clustering to particular solutions of nonlinear evolution equations like Bose-Einstein condensation or solitons.

Firstly, the experimental evidence of cluster decay as spontaneous emission of carbon, neon, magnesium and silicon from heavy nuclei, indicates a large enhancement of such clusters on the nuclear surface. By considering nonlinear terms in the hydrodynamics of the liquid drop model for the nucleus, it was inferred that KdV solitary waves could exist on the surface of nuclei, and explain cluster decay as a large amplitude collective excitation \cite{first,x21,gherghescu}. Nevertheless, in order to reproduce the experimental spectroscopic factors for alpha and cluster decays with such a nonlinear integrable model defined on the nuclear surface it was necessary to add shell corrections. Thus one obtained  a coexistence model consisting of the usual shell model and a cluster-like model, leading to a minimum in the total potential energy  degenerated with the ground state minimum.

Secondly, $\alpha-$like states were detected for many light to heavy nuclei. The alpha clustering in the nuclear structures, and clustering models have a long history, but in the last decade  a rapid development successfully  explained the structure of
many states in light to heavy nuclei, especially in $n \cdot \alpha$ nuclei \cite{2018alpha}.

Thirdly, a moment of inertia anomaly was emphasized in the rotational bands of resonance elastic scattering measurements. By plotting the mean weighted values with the reduced widths of the experimental energy levels vs. $J(J+1)$ for such experiments one can obtain a value for the moment of inertia of the system. By comparing this value with  the theoretic moment of inertia of an alpha particle plus the daughter nucleus rotating together at touching distance we have a discrepancy: the experiment provides smaller values by a factor of at least 2 than the rigid rotor moment of inertia \cite{neon,finland,2011alpha}. For example, in \cite{2011alpha} the measured moment of inertia for the elastic scattering $\alpha + ^{36}$Ar $\rightarrow ^{40}$Ca was $\mathcal{I}=3.8\pm 0.3 \hbar^2/$MeV, while alpha particle orbiting a non-interacting $^{36}$Ar core would have a $\mathcal{I}=9.3 \hbar^2/$MeV moment of inertia. Moreover, this larger theoretical value results also from calculations of the strongest superdeformed bands in $^{40}$Ca ($4p-4h$ and $8p-8h$ excitations).
It appears that the geometric configuration of a cluster orbiting around a daughter nucleus is a little more complicated than a rigid rotor.

The fourth observation is related to various aspects of collective motion in nuclei. 
One of the collective motion degree of freedom is caused by  the spontaneous symmetry 
breaking of rotational invariance due to the alpha clustering.
The collective motion related to  cluster condensation, or superfluidity in nuclei,
has been paid attention in
the frame of the many-body theory in the last decades.
In \cite{2018bec} it was shown that collective states of the zero mode operators
are new-type soft modes due to Bose-Einstein condensate of alpha clusters. 

From the Bose-Einstein phenomenological Hamiltonian of a number of alpha particles 
trapped by an external potential it results a Gross-Pitaevskii equation (G-P) for 
the nuclear condensate component of the field operator. The solution of the G-P equation  is the order parameter of the phase transition from the Wigner phase to the 
Nambu-Goldstone phase is a superfluid amplitude, square of the modulus
of which is the superfluid density distribution \cite{2018bec}, which ultimately describes the nuclear shape. The G-P equation is in the same hierarchy as the NLS equation just having in addition the potential trap linear term. Consequently, it is natural for the  G-P equation to have solitary waves solutions as shown for example in \cite{solib,BEC1,BEC2}.

This final observation supports the NLS droplet model, since both G-P and NLS equations have similar type of solitons, and can be equally used to explain some exotic nuclear shapes. The NLS model for quantum droplets presented here has the advantage of being already quantized, so it does not request shell corrections when  applied in nuclear models.

In all experimental comparison we use only three fitting parameters, namely $c$ (the shape  parameter), $N$ and $E_0$, the last one being a multiplicative re-scaling of the energy spectra in Eq. (\ref{eq.bethe1}). The expression of the total energy is the sum between the scaled Bethe spectrum and a rigid rotation  $E_{exp}=  E_0 E_N (c)+\hbar^2 J(J+1) / (2 \mathcal{I})$.

In Fig. \ref{fig1} we present a comparison between experimental spectra \cite{neon}, of resonant states obtained in the collision of $\alpha + ^{20}$Ne$\rightarrow  ^{24}$Mg  generating bound $\alpha-$cluster states $J=0^{+}$ to $J=7^{-}$ and the theoretical Bethe spectra Eq. (\ref{eq.bethe1}) for $N=3$ and $4$ from our model. We present the results  for $c$ providing the best fit with the experiments. We noticed that  odd-parity states are associated to larger values for $c>c_{crit.}$, while  even-parity states are associated with relative smaller values for $c$.

In both Figs. \ref{fig2} and \ref{fig3} we present medium heavy nuclei where the  $\alpha-$daughter quasimolecular rotational bands are manifest. From the slope of the mean positions of the rotational bands we obtain the value of the moment of inertia: $\mathcal{I}=3.3 \pm 0.2 \hbar^2 /$MeV for $^{28}$Si and $\mathcal{I}=3.8 \pm 0.3 \hbar^2 /$MeV for $^{36}$Ar which are almost half of the values for rigid rotor configuration for the given masses and radii.

In Fig. \ref{fig2} we present a comparison between experimental energy spectra, \cite{finland}, of resonant states for the elastic collision of $\alpha-$particles on $^{28}$Si targets and the theoretical Bethe spectra Eq. (\ref{eq.bethe1}) for $N=3$ and $4$ and with the parameter $c$ chose to provide the best fit with experiments. We remark that the odd angular momentum states are associated with larger values of $c$. Also, the states with $J=6,7$ have the best theoretical fit for rapidity $N=3$ while all the other states are related to $N=4$.

In Fig. \ref{fig3} we present  another set of experimental energy spectra, \cite{2011alpha}, for the resonant states during collisions of $\alpha-$particles on $^{36}$Ar targets, in comparison with the theoretical Bethe spectra Eq. (\ref{eq.bethe1}). One notices more theoretical energy  levels than the experimental spectra, but this may be explained by  the limitations in data availability. There are two observations in favor of the NLS droplet model support. On one hand, experimental states with $J=5$ present an anomaly of having very few resonances, while the NLS model also predicts a very sparse density of states when fitted at $N=3$ with energies measured at this $J$ value. On the other hand, all odd parity nuclear states from experiment fit the best at $c \sim 1$ which represents solitons orbiting the equator of the droplet, so high values of angular momentum, while even parity states fit rather $c \ll 1$ which corresponds to solitons orbiting around the poles of the droplet, hence low angular momentum.

In Fig. \ref{fig4} we present two types of comparisons. In the first column we plot the experimental energies of  bound $\alpha-$cluster resonances in $^{12}$C together with the theoretical Bethe spectra, Eq. (\ref{eq.bethe1}). We notice that, even for the best fit,  our model generates the first three energy levels below the Hoyle state. In the  next four columns we fit our theoretical spectra with a field theoretical super-fluid cluster model \cite{2018bec}. In this model, spontaneous symmetry breaks the global Wigner phase in a finite number $n$ of $\alpha$ clusters, a Bose-Einstein condensation process. We compare our Bethe spectra for $N=4$ with  spectra resulting from condensation to $n=6$ ($^{24}$Mg) and $n=8$ ($^{32}$S) $\alpha-$clusters, for two different available condensation rates of $70\%$ and $100\%$ calculated in \cite{2018bec}. In these cases, the comparison results in a good match between the two models above the Hoyle state (here $0$ MeV).

Fig. \ref{fig5} represents a wide-range comparison for the energies of bound $\alpha-$cluster states: from light $^{20}$Ne, to medium $^{44}$Ti, to superheavy $^{212}$Po nuclei, with the theoretical Bethe spectra Eq. (\ref{eq.bethe1}).  While we do not have a perfect match for each spectral line,  the structure and density of spectral lines is matched surprisingly well for all masses, energies and angular momenta, in spite of the fact that we have only two free parameters $E_0, c$, plus the choice of which value for rapidity $N$ and which spectral tower to use. We note that lighter nuclei are fitted better by smaller rapidity  ($N=3$) and larger form parameter ($c\sim 5$), while heavier nuclei are fitted better by larger rapidity  ($N=4$) and smaller values for the form parameter ($c \ll 1$), while obeying the same rule: the larger the atomic mass or number of alpha clusters, the larger $c$ values fit better. Moreover, these experimental spectra were used for comparisons with the predictions of the quartet model, \cite{2018NeTi}, or the density-dependent cluster model plus the two-potential approach for heavy nuclei \cite{2018TeXe}. The intrinsic wave function of the quartet acquires  cluster configuration,  when it orbits a radius above the core nucleus, similar to the surface formation of the soliton solution, Eq. (\ref{eqsolu}), in the defocusing NLS nuclear shape model.

\section{Conclusions}
In this paper we have developed an asymptotic description of azimuthal envelope solitons on spherical liquid layers as solutions of defocusing nonlinear Schr\"{o}dinger equation. The quantum dynamics is analyzed using the algebraic Bethe ansatz, showing a spectrum of a rigid rotor for weak nonlinearity (measured by the coefficient of nonlinear term in the NLS equation) and an oscillatory-type spectrum for strong nonlinearity. 
On the other hand the approximation used to get the NLS equation needs to be improved to include the evolution in the polar coordinate as well. We expect that fully localized lump-type solutions to move on the surface of the spherical liquid. However because of the spherical geometry it is almost sure that such an equation will be a non-autonomous one and only numerical simulations will show interesting facts.
In order to validate the model, we compare its theoretical predictions in terms of energy excitations with a large set of nuclear experimental data of energies of cluster resonant states or elastic collisions, for a variety of atomic masses from light to superheavy nuclei, and for the corresponding angular momenta and parities. The NLS models seem to offer a good fit with the structure of experimental nuclear spectra. This NLS droplet model is intended to elaborate on a number of theoretical questions, in an effort to be  a useful complement to phenomenological and microscopic models, and help deepening our understanding on clustering phenomena and decays across the chart of nuclides.

\end{document}